\documentclass[12pt]{article}
\usepackage{amsmath,amssymb,amsthm,amsxtra,overpic,bbm,bm,epsfig,ulem,multirow}
\usepackage{color,cite}
\usepackage{epstopdf}

\def\thefootnote{\fnsymbol{footnote}}

\usepackage{array}
\usepackage{textcomp}

\begin{document}

\vspace{0.2cm}

\begin{center}
{\Large\bf Ab initio calculations of reactor antineutrino fluxes with exact lepton wave functions}
\end{center}

\vspace{0.2cm}

\begin{center}
{\bf Dong-Liang Fang$^{a,b}$}\footnote{Email: dlfang@impcas.ac.cn}\\
{\small $^a$ Institute of Modern Physics, Chinese Academy of sciences, Lanzhou, 730000, China}\\
{\small $^b$ School of Nuclear Science and Technology, University of Chinese Academy of Sciences, Beijing 100049, China}\\

{\bf Yu-Feng Li$^{c,d}$}\footnote{Email: liyufeng@ihep.ac.cn (corresponding author)}
and {\bf Di Zhang$^{c,d}$}\footnote{Email: zhangdi@ihep.ac.cn (corresponding author)} \\
{\small $^c$ Institute of High Energy Physics, Chinese Academy of Sciences, Beijing 100049, China}\\
{\small $^d$ School of Physical Sciences, University of Chinese Academy of Sciences, Beijing 100049, China}\\
\end{center}

\vspace{1.5cm}

\begin{abstract}
New \textit{ab initio} calculations of the isotopic reactor antineutrino fluxes are provided with exact numerical calculations of the
lepton wave functions, assuming all the decay branches are allowed GT transitions. We illustrate that the analytical
Fermi function and finite size effect each could have the largest spectral deviation of $\mathcal{O}(10\%)$, whereas the effect of their combination
could result in spectral deviations at the level of 5\%---10\%.
Meanwhile, we also find that several forms of the extended charge distributions have negligible effects on the spectral variation.
Using the state-of-the-art nuclear databases, compared to usual \textit{ab initio} calculations using the analytical single beta decay spectrum,
our new calculation shows sizable but opposite
spectral deviations at the level of 2\%---4\% for the cumulative antineutrino and electron energy spectra which may partially contribute to the observed
spectral excess in the high energy antineutrino range.
Finally we observe that the {bias} of analytical beta decay spectrum approximation is rather universal for all the four fissionable isotopes.
\end{abstract}

\newpage

\def\thefootnote{\arabic{footnote}}
\setcounter{footnote}{0}

\section{Introduction}

Electron antineutrinos from nuclear reactors have been widely used to study the fundamental properties of massive neutrinos~\cite{Tanabashi:2018oca}.
{Reactor antineutrinos are the products of the beta decay of fission fragments} associated with four main fissionable isotopes
${}^{235}\text{U}$,
${}^{238}\text{U}$,
${}^{239}\text{Pu}$, and
${}^{241}\text{Pu}$.
Predicting the reactor antineutrino flux and spectrum is always an important prerequisite for the reactor antineutrino experiments~\cite{Hayes:2016qnu,
Davis:1979gg,Avignone:1980qg,Klapdor:1982zz,Klapdor:1982sf,Mueller:2011nm,Fallot:2012jv,Hayes:2013wra,Dwyer:2014eka,
Hayes:2015yka,Littlejohn:2018hqm,Yoshida:2018zga,Estienne:2019ujo,VonFeilitzsch:1982jw,Schreckenbach:1985ep,Hahn:1989zr,Haag:2013raa,
Vogel:2007du,Huber:2011wv,Hayen:2019ieh,Li:2019quv,Hayen:2019eop}.
There are two different methods to obtain the theoretical calculation of the reactor antineutrino flux and spectrum.
The first one employs the \textit{ab initio} method~\cite{Davis:1979gg,Avignone:1980qg,Klapdor:1982zz,Klapdor:1982sf,
Mueller:2011nm,Fallot:2012jv,Hayes:2013wra,Dwyer:2014eka,
Hayes:2015yka,Littlejohn:2018hqm,Yoshida:2018zga,Estienne:2019ujo}
by a direct summation of all the beta decay branches using the available information from the latest nuclear databases.
The other method uses an effective conversion
procedure of virtual branches~\cite{VonFeilitzsch:1982jw,Schreckenbach:1985ep,Hahn:1989zr,Haag:2013raa,
Vogel:2007du,Huber:2011wv,Hayen:2019ieh,Li:2019quv,Hayen:2019eop}
based on the measurements of the integral electron energy spectra of the fissionable isotopes.

During the reactor fuels burning, more than six thousand beta decay branches contribute to the antineutrinos produced in each fissionable isotope.
In the first method of the \textit{ab initio} calculation, the fission yield, the endpoints and branching ratios of each fission fragment can
be accessed from the nuclear databases. The reactor antineutrino flux can be obtained by a direct summation of all the beta decay
branches using an analytical description of the single beta decay spectrum~\cite{Davis:1979gg,Avignone:1980qg,Klapdor:1982zz,Klapdor:1982sf,
Mueller:2011nm,Fallot:2012jv,Hayes:2013wra,Dwyer:2014eka,
Hayes:2015yka,Littlejohn:2018hqm,Yoshida:2018zga,Estienne:2019ujo}.
Therefore, the {accuracy} depends on the uncertainties of the fission yields and the beta decay information.
The second effective conversion method has been developed using the corresponding electron
spectrum associated with the thermal neutron induced fission of
${}^{235}\text{U}$, ${}^{239}\text{Pu}$, and
${}^{241}\text{Pu}$ at ILL, Grenoble, France in 1980s~\cite{VonFeilitzsch:1982jw,Schreckenbach:1985ep,Hahn:1989zr}, and the fast neutron induced fission of ${}^{238}\text{U}$ at FRMII in Garching, Germany in 2011~\cite{Haag:2013raa}.
The antineutrino flux is calculated by assuming dozens of
virtual branches and fitting the electron spectrum of the fission isotope to be consistent with the measurement.

However, the experimental measurements of the total inverse-beta-decay rate and energy spectrum have shown anomalous results compared
to the theoretical predictions~\cite{Mention:2011rk,Giunti:2017yid,Giunti:2019qlt,Berryman:2019hme,Giunti:2019fcj,Seo:2014xei,An:2015nua,Adey:2019ywk,Abe:2015rcp,Ko:2016owz}.
On the one hand, there is a 6\% deficit in the total rate measurement when one employs the new evaluation of the reactor antineutrino flux~\cite{Mention:2011rk,Giunti:2017yid,Giunti:2019qlt,Berryman:2019hme,Giunti:2019fcj}. On the other hand, according to the latest reactor experiments,
a bump-like structure of event excess near the region of 5 MeV for the observed positron energy has been consistently observed~\cite{Seo:2014xei,An:2015nua,Adey:2019ywk,Abe:2015rcp,Ko:2016owz}.
These reactor anomalies of rate and spectral measurements have challenged the validity of the theoretical calculation of the
reactor antineutrino flux and triggered intensive studies on how to make an accurate prediction of the reactor
antineutrinos from the fission isotopes~\cite{Fallot:2012jv,Hayes:2013wra,Dwyer:2014eka,
Hayes:2015yka,Littlejohn:2018hqm,Yoshida:2018zga,Estienne:2019ujo,Hayen:2019ieh,Li:2019quv,Hayen:2019eop}.

In the \textit{ab initio} method, the fission yields have been evaluated by different nuclear databases,
{which constitute an important contributor to the overall uncertainties of summation calculations.}
In addition, the beta decay information including the branching ratios, endpoints and the quantum numbers of the initial and final states
are not always known for the existing fragments. There is a well known problem of the Pandemonium effect~\cite{Hardy:1977suw}
 for some nuclear beta decay data that has been proved to play important roles in the reactor antineutrino flux calculations~\cite{Fallot:2012jv,Estienne:2019ujo}.
Moreover, although theory of the analytical description of single beta decay spectrum, based on the series expansions of the nuclear charge and radius,
as well as the electron and antineutrino energies, has been developed since 1960s~\cite{BB-Book},
the accuracy and validity of the analytical description on reactor antineutrino predictions, which have beta decay branches with large nuclear charge and/or
high endpoint energies, have not been carefully studied until recently~\cite{Fang:2019qcu}.

In this work, by assuming all the decay branches are allowed GT transitions
we employ a new \textit{ab initio} calculations of the isotopic reactor antineutrino fluxes using the exact calculations of the
numerical lepton wave functions.
We make a systematic test on the reliability and accuracy of the analytical description of the single beta decay spectrum,
and observe that depending on the endpoint energies and nuclear charges,
the analytical Fermi function and finite size effect could have the largest spectral deviation of $\mathcal{O}(10\%)$,
whereas the effect of their combination may result in spectral deviations at the level of 5\%---10\%.
{Compared to the usual \textit{ab initio} calculations using the analytical single beta decay spectrum,
our new calculation of reactor antineutrinos using exact lepton wave functions} shows sizable but opposite
spectral deviations at the level of 2\%---4\% and the {bias} of analytical beta decay spectrum approximation is rather universal for all the four fissionable isotopes.
{Note that further refinement of the numerical calculations of reactor antineutrino spectra is necessary, including the latest Pandemonium free data (as done in Ref.~\cite{Estienne:2019ujo}) and other spectral corrections of beta decays.}

This work is organized as follows. In Sec.~2 we review the description of beta decays theory and discuss how to make the {approximations}
towards the usual analytical calculations.
Then we investigate the dependence of these approximation on the beta decay endpoint energies and nuclear charges in Sec.~3.
In Sec.~4 the new exact calculations of the single beta decay spectrum are applied to the four reactor fission isotopes,
${}^{235}\text{U}$, ${}^{238}\text{U}$, ${}^{239}\text{Pu}$, and
${}^{241}\text{Pu}$. Finally the concluding remarks are presented in Sec.~5.

\section{Theoretical description of the beta decay}

In this section, we want to review the theoretical description of the beta decay and discuss how to make the approximation towards
the standard analytical calculations.

In the Standard Model, the nuclear beta decay is mediated by the charged $SU(2)_L$ gauge boson $W$
whose mass is generated by the spontaneous symmetry breaking. For the nuclear decay with energy at the range of several MeV,
the weak interaction can be effectively described by the contact interactions of hadron currents and lepton currents following conventions in~\cite{Fermi:1934hr}:
\begin{eqnarray}
{\mathcal{H}} &=&\frac{G_\beta}{\sqrt{2}}[\overline{\psi}_n(\vec{r}_1) (1+g_A\gamma_5) \psi_p(\vec{r}_1\mathbf{})]\nonumber \\
&\times& \delta(\vec{r}_1-\vec{r}_2) [\overline{\psi}_e(\vec{r}_2)(1+\gamma_5)\psi_{\nu}(\vec{r}_2)]\,,
\end{eqnarray}
where $G_\beta=G_{\rm F}\cos\theta_{\rm C}$, with $G_{\rm F}$ being the Fermi constant and $\theta_{\rm C}$ being the Cabbibo angle.
${\psi}_n$, ${\psi}_p$ ${\psi}_e$, and ${\psi}_{\nu}$ are the wave functions of the neutron, proton, electron and antineutrino respectively.

The bounded nucleus system usually have the spherical symmetry, thus in order to describe such system, the spherical coordinate is the most appropriate one.
Therefore, the contact interaction can be decomposed to summations of various angular transfer interactions (We choose the definitions of spherical harmonics as in Appendix VIII of~\cite{Wei61}):
\begin{eqnarray}
\delta(\vec{r}_1-\vec{r}_2)=\frac{\delta(r_1-r_2)}{r_1r_2}\sum_{LM}(-1)^{M} Y_{L}^{M}(\hat{r}_1) Y_{L}^{-M}(\hat{r}_2)
\end{eqnarray}
where $Y_{L}^{M}$ is the spherical harmonic function, $\hat{r}_{i}=\vec{r}_{i}/r_{i}$ with $r_{i}=|\vec{r}_{i}|$ ($i=1,2$).
Thus the Hamiltonian with the definite angular momentum and parity can be written as:
\begin{eqnarray}
H = \int \mathcal{H}\;d^3\vec{r}_{1}d^3\vec{r}_{2} &=&\sum_{KLsM}\int^{\infty}_{0} r^2 dr \int d\Omega_N \overline{\psi}_p(r,\Omega_N)(1+g_A\gamma_5)T_{KLs}^{M} \psi_n(r,\Omega_N) \nonumber \\
&\times&\int d\Omega_L \overline{\psi}_e(r,\Omega_L) (1+\gamma_5){T_{KLs}^{-M}} \psi_\nu(r,\Omega_L)\,,
\end{eqnarray}
where $\Omega_N$ and $\Omega_L$ are the solid angle coordinates of the nucleus and lepton systems respectively.
$T_{KLs}^{M}$ are the irreducible multi-pole tensor operators of rank $K$ and in general can be expressed as the products of spherical harmonics
and Dirac matrices, whose expressions can be found in Refs.~\cite{Wei61,BBC77}.
$K$, $L$ and $s$ are the quantum numbers of the total, orbital and spin angular momenta of the mediating $W$ boson, respectively, and $M$ is the
magnetic quantum number of the total angular momentum. 


For a definite weak transition with known spin and parity of the initial and final nuclei, only those transition operators with certain selection rules are relevant.
A naive estimation shows that the electron and antineutrino wave functions are proportional to $({p_{e}}R)^{l_{e}}$ and $({p_{\nu}}R)^{l_{\nu}}$ respectively
when ${p_{e}}R$ and ${p_{\nu}}R$ are small, where $R$ is the nuclear radius, $p_{e}$ ($p_{\nu}$) and $l_{e}$ ($l_{\nu}$) are the momentum and orbital angular momentum
of the outgoing electron (antineutrino) respectively.
Therefore the larger angular momentum transfer would imply the smaller beta decay strength.
This suggests that the largest weak decay happens for transitions with $l_{e}=0$ and $l_{\nu}=0$,
which, corresponding to the selection rules of $\Delta J^\pi=0^+$ and $1^+$, are usually defined
as the allowed Fermi and allowed Gamow-Teller (GT) transitions, respectively.
Transitions with the larger angular momentum transfer are called the forbidden decays.
{In this work we assume all transitions associated with the fission fragments are allowed GT transitions and temporarily neglect the effects of forbidden
decays}~\footnote{{In the latest studies~\cite{Hayes:2013wra,Hayen:2019ieh,Li:2019quv,Hayen:2019eop},
it is shown that the first forbidden transitions may be important to explain the anomalous
results of the reactor rate and spectral measurements. The numerical calculations of spectral variations in forbidden decays will be reported in a future separated study.}}.

For these allowed GT decays, we have the leading contribution of $K=1$, $L=0$ and $s=1$, and the differential decay width can be calculated as~\cite{Wei61,BBC77}:
\begin{eqnarray} 
\frac{d\lambda_{\rm GT}}{d E_{e}}&=&\frac{G_\beta^2 }{( 2 J_i+1)2\pi^3} \sum_{\kappa_e, \kappa_\nu>0} p_{e} E_{e} (Q-E_{e})^2  \nonumber \\
&\times& \left\{ \int  \frac{{{g_A}}}{\sqrt{3}}\langle\langle J^{\pi_{f}}_f || \sigma || J_i^{\pi_{i}} \rangle\rangle F_{\kappa_e,\kappa_\nu}(p_{e},Q-E_{e},r) r^2 dr \right\}^2\,,
\end{eqnarray}
where $Q=M_{i}-M_{f}$ is the endpoint energy, $M_{i}$ ($J_{i}$, $\pi_{i}$) and $M_{f}$ ($J_{f}$, $\pi_{f}$) are the mass (spin, parity) of
the initial and final nucleus respectively, $p_{e}$ and $E_{e}$ are the momentum and energy of the outgoing electron.
$\langle\langle J^{\pi_{f}}_f || \sigma || J_i^{\pi_{i}} \rangle\rangle $ is the nuclear matrix element of the GT transition,
where the double bras and kets refer to integrations over the solid angle only.
\begin{eqnarray}
F_{\kappa_e,\kappa_\nu}(p_{e},Q-E_{e},r) = \sqrt{4\pi}\langle\langle \phi_{\kappa_e}(Z) || (1+\gamma_{5})T_{101}^{0} || \phi_{\kappa_{\nu}} \rangle\rangle\,,
\end{eqnarray}
is the lepton matrix element, where $\phi_{\kappa_e}(Z)$ and $\phi_{\kappa_{\nu}}$ are radial wave functions of the electron and antineutrino respectively,
where $Z$ is the nuclear charge, $\kappa_e$ and $\kappa_\nu$ are defined as
\begin{eqnarray}
\kappa_{i}=\left\{ \begin{matrix} l_{i} \quad\quad &{\rm for}& \quad\quad  j_{i}=l_{i}-\frac{1}{2}
\\ -(l_{i}+1) \quad\quad &{\rm for}& \quad\quad j_{i}=l_{i}+\frac{1}{2} \end{matrix} \right.\,,
\end{eqnarray}
where $i=(e,\nu)$, $j_{e}$ and $j_{\nu}$ are the total angular momenta of the electron and antineutrino respectively.

To proceed, one needs to calculate each of the nuclear and lepton wave functions, however, the nuclear wave functions are rather complicated
and rely on the nuclear structure models. In order to factorize the nuclear and lepton matrix elements, a normal assumption is to treat the
lepton radial wave functions as constants at the surface of the nuclear distribution, namely, the surface approximation (SA) taken as
\begin{eqnarray}
F_{\kappa_e,\kappa_\nu}(p_{e},Q-E_{e},r)\simeq F_{\kappa_e,\kappa_\nu}(p_{e},Q-E_{e},R)\,,
\end{eqnarray}
where $R$ is the nuclear radius. This will simplify the differential decay width as
\begin{eqnarray}
	\label{eqSA}
\frac{d\lambda_{\rm GT}}{d E_{e}}&=&\frac{G_\beta^2 }{( 2 J_i+1)6\pi^3} \left\{ \int {{g_A}} \langle\langle J^{\pi'}_f || \sigma || J_i^{\pi} \rangle\rangle  r^2 dr \right\}^2 \nonumber \\
&\times&    \sum_{\kappa_e, \kappa_\nu>0}  p_{e} E_{e} (Q-E_{e})^2 F_{\kappa_e,\kappa_\nu}^2(p_{e},Q-E_{e},R)\,,
\end{eqnarray}
where the GT transition strength is defined as:
\begin{eqnarray}
B({\rm GT})=\left\{ \int  \langle\langle J^{\pi'}_f || \sigma || J_i^{\pi} \rangle\rangle  r^2 dr \right\}^2=|\langle f ||\sigma ||i \rangle|^2\,.
\end{eqnarray}
Notice that $F_{\kappa_e,\kappa_\nu}$ can be calculated using the antineutrino radial wave function $j_{l_{\nu}}$ of the spherical Bessel function,
and the electron radial wave functions $f_{\kappa_{e}}$, $g_{\kappa_{e}}$, which can be derived by using the following evolution equations~\cite{Wei61,BBC77}:
\begin{eqnarray}
\frac{{d}\,f_{\kappa_{e}}}{{d}\,r}&=&+\frac{\kappa_{e}-1}{r}f_{\kappa_{e}}-[E_{e}-m_{e}-V(r)]g_{\kappa_{e}}\,,\\
\frac{{d}\,g_{\kappa_{e}}}{{d}\,r}&=&-\frac{\kappa_{e}-1}{r}g_{\kappa_{e}}+[E_{e}+m_{e}-V(r)]f_{\kappa_{e}}\,,
\end{eqnarray}
where $V(r)$ is the Coulomb potential of centrally distributed nuclear charges.



The next step is to evaluate the lepton wave functions under the Coulomb potential, which can be calculated numerically,
and in some special cases can be analytically expressed~\cite{Fermi:1934hr}.
In this work, we employ the {\it Radial} package~\cite{Fang:2019qcu,SFW95} and use Eq.~(8) to calculate the exact numerical solutions for the radial electron
wave function for any given nuclear charge distributions.
In contrast, to obtain the well known analytical Fermi function and finite size effect, several additional approximations have to be implemented:
\begin{itemize}
\item \textit{The neutrino long-wave approximation} ($\nu$LA).
Antineutrino produced from the nuclear beta decay can be described by the plane wave under the multi-pole expansions. The neutrino long-wave approximation is defined by taking the $s$-wave neutrino component to be unity and neglecting all other components:
 \begin{eqnarray}
  j_{0}(p_{\nu}r)=1\,,\quad{\rm and}\quad j_{l_{\nu}}(p_{\nu}r)=0\quad{\rm with}\quad l_{\nu}\neq0\,,
\end{eqnarray}
where $j_{l_{\nu}}(p_{\nu}r)$ is the spherical Bessel function with the angular momentum $l_{\nu}$. Then the differential decay width will be reduced to
\begin{eqnarray}
	\label{eqLA}
\frac{d\lambda_{\rm GT}}{d E_{e}}&=&\frac{G_\beta^2}{( 2 J_i+1)2\pi^3} g^{2}_A  B({\rm GT})
p_{e} E_{e} (Q-E_{e})^2 [g^{2}_{-1}(p_{e}R)+f^{2}_{1}(p_{e}R)]\,,
\end{eqnarray}
where $f_{1}(p_{e}R)$ and $g_{-1}(p_{e}R)$ are radial electron wave functions at the radius $R$ with $\kappa_e=1$ and $\kappa_e=-1$ respectively.

\item \textit{Further series expansions} up to {$\mathcal{O}(\alpha Z)$, $\mathcal{O}(p_{e}R)$ and $\mathcal{O}(m_{e}R)$}.
For the case of the point Coulomb potential, $f_{1}(p_{e}R)$ and $g_{-1}(p_{e}R)$ can be analytically calculated and
after the series expansions up to the orders of {$\mathcal{O}(\alpha Z)$, $\mathcal{O}(p_{e}R)$ and $\mathcal{O}(m_{e}R)$},
one can derive the following expression for the Fermi Function~\cite{Wei61}:
\begin{eqnarray}
{g_{-1}^2(p_{e}R)+f_{1}^2(p_{e}R)\approx F(Z,E_{e}) =2(1+\gamma)(2p_{e}R)^{2(\gamma-1)} e^{\pi y} \frac{|\Gamma(\gamma+iy)|^2}{|\Gamma(2\gamma+1)|^2}} \;,
\end{eqnarray}
{where $\gamma = \sqrt{1-(\alpha Z)^2}$ and $y = \alpha Z E_{e}/p_{e}$ with $\alpha$ being the fine structure constant.} 

\item \textit{The finite size effect}. For the extended charge distributions in the nucleus, such as the uniform, the Gaussian or the Fermi charge distribution,
besides the Fermi function of the point charge potential, there will additional terms to account for the finite size effect.
There are many different analytical calculations of finite size corrections based on different methods and assumptions~\cite{Wilkinson,Wang:2016rqh}.
In this work, in order to compare with the exact numerical calculations, we choose the analytical finite size effect~\footnote{{In order to make an apple-to-apple comparison, we have neglected the weak finite size term $C(Z,W)$~\cite{Wilkinson}, which is induced by the distribution of nuclear weak charge, and depends on the nuclear structure calculation.}}
obtained by Wilkinson~\cite{Wilkinson}
using a uniform charge distribution within the nuclear radius $R$:
\begin{eqnarray}
\delta^{}_{FS} &=& \frac{2}{\gamma+1} L_0 - 1 \;,
\end{eqnarray}
with
\begin{eqnarray}
L_0 &=& 1 + \frac{13}{60} (\alpha Z)^2 - \frac{41-26\gamma}{15(2\gamma-1)}\alpha Z WR - \frac{\gamma(17-2\gamma)}{30(2\gamma-1)} \frac{\alpha Z R}{W}
\nonumber
\\
& & + a_{-1} \frac{R}{W} + \sum_{n=0}^{5} a_n(WR)^n + 0.41(R-0.0164)(\alpha Z)^{4.5} \;,
\end{eqnarray}
where $W=E_{e}/m_{e}$, $R = 0.0029A^{1/3} + 0.0063A^{-1/3} -0.017A^{-1}$ in the unit of $m_{e}$ with $A$ being the mass number of the nucleus.
 The definition of $a_n$ ($n=-1,0,\cdots,5$) can be found in Ref.~\cite{Wilkinson}.
\end{itemize}

After all the above simplification, one can finally arrive at the famous analytical expression for the allowed GT transition:
\begin{eqnarray}
	\label{eqFermi}
\frac{d\lambda_{\rm GT}}{d E_{e}}&=&\frac{G_\beta^2}{( 2 J_i+1)2\pi^3} g^{2}_A  B({\rm GT})
p_{e} E_{e} (Q-E_{e})^2F(Z,E_{e})(1+\delta^{}_{FS})\,.
\end{eqnarray}
The antineutrino energy spectrum can be obtained by the direct replacement $E_{e} \rightarrow Q - E_{\nu}$.
Notice that there exist some additional corrections~\cite{Huber:2011wv} in Eq.~(17), such as those of the screening effect, weak magnetism and radiative corrections,
but to make our study easier, we temporarily neglect their contributions to the beta decay spectrum in this work.

\section{Single beta spectrum investigation}

Before calculating reactor antineutrino spectra with the \textit{ab initio} method, we want to evaluate the accuracy of the analytical expressions
derived in the previous section by comparing with the exact numerical solutions of the electron radial wave functions.
Three different approximation effects, including the $\nu$LA, the analytical Fermi function and the finite size effect,
will be tested by selecting four representative beta decay branches with different endpoint energies and nuclear charges, which are summarized in Table~1.
\begin{table}[h]
	\caption{Representative beta decay branches to illustrate the effects of the $\nu$LA, analytical
Fermi function and finite size correction on the individual spectrum.}
	\centering
	\vspace{8pt}
	\setlength{\tabcolsep}{12mm}{
\begin{tabular}{cccc}
	\hline
	\hline
	& $Z$  & $A$ & $ Q\;[{\rm MeV}]$ \\
	\hline
	\multirow{2}{*}{$^{142}{\rm Cs}$}&	\multirow{2}{*}{55} & 	\multirow{2}{*}{142} & 3.450 \\ \cline{4-4}
	&  & & 7.819 \\
	\hline
	\multirow{2}{*}{$^{92}{\rm Rb}$}&	\multirow{2}{*}{37} & 	\multirow{2}{*}{92} & 3.552 \\ \cline{4-4}
	&  & & 7.791 \\
	\hline
	\hline
\end{tabular}}
\end{table}

Let us first consider the point charge distribution of the Coulomb potential and test the validity for the approximations of
the $\nu$LA and the analytical Fermi function. The numerical calculations according to the Eq.~(8) will be used as our benchmark to make the comparison,
in which the GT transition strength is normalized to reproduce the experimental decay lifetime.
Results of the $\nu$LA and the Fermi function are calculated using Eq.~(13) and Eq.~(14) respectively.
Comparisons between the numerical calculations and analytical approximations of four representative decay branches are illustrated in Fig.~1,
where the spectral ratios are defined as numerical calculations of the point charge distribution (PD) with (red, {see Eq~(\ref{eqLA})}) and without (blue, {see Eq~(\ref{eqSA})}) the $\nu$LA to the analytical Fermi function approximation {(FFA, {see Eq~(\ref{eqFermi}) without $\delta_{FS}$})} for each beta decay branch. The solid and dashed lines are for the electron and antineutrino spectra respectively. Several comments are provided as follows:
\begin{figure}[t]
	\centering
	\includegraphics[width=1\linewidth]{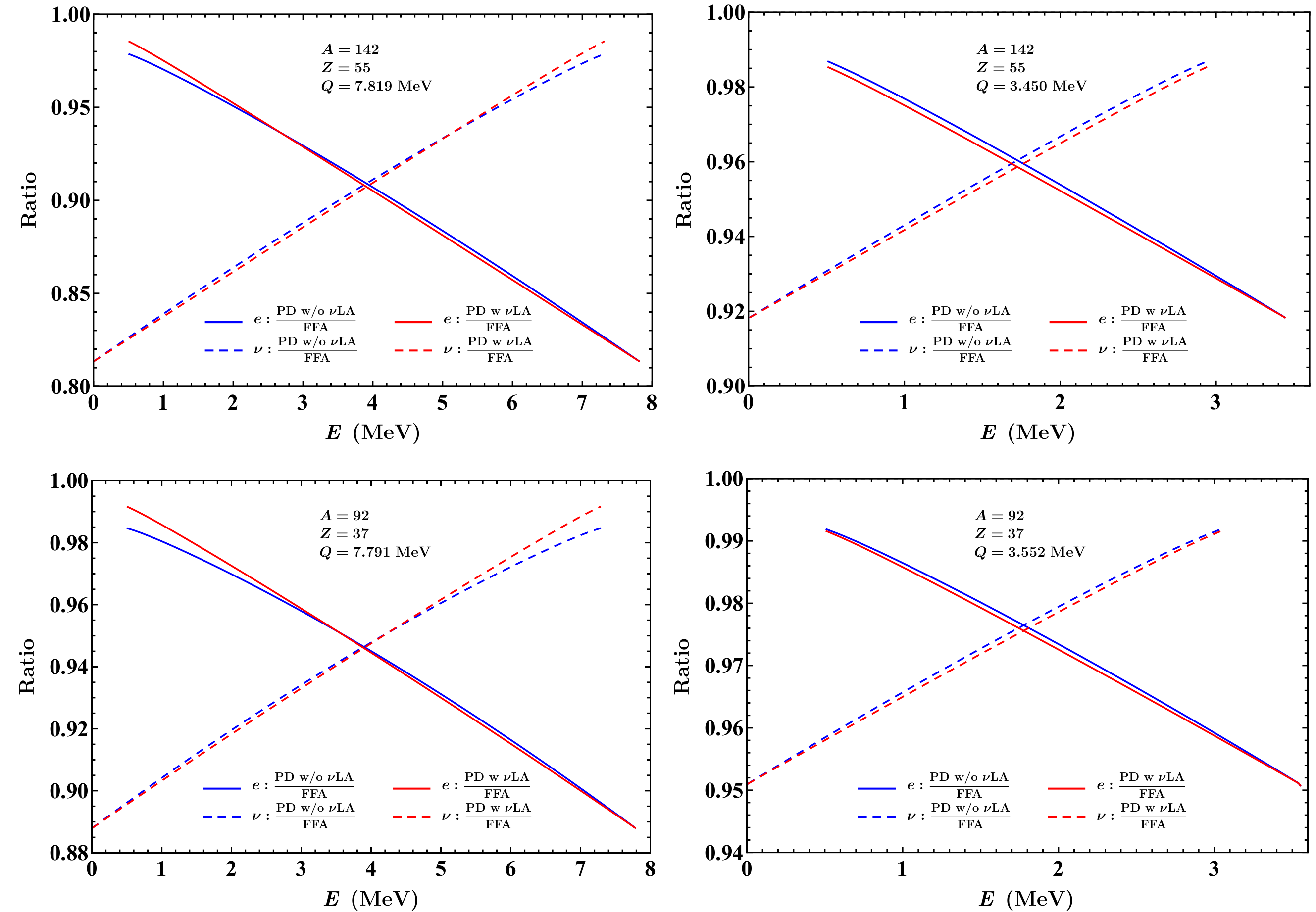}
	\caption{Spectral ratios of the numerical calculations for the point charge distribution (PD) with (red, {see Eq~(\ref{eqLA})}) and without (blue, {see Eq~(\ref{eqSA})}) the $\nu$LA to the analytical Fermi function approximation {(FFA, {see Eq~(\ref{eqFermi}) without $\delta_{FS}$})} for each beta decay branch. The solid and dashed lines are for the electron and antineutrino energy spectra respectively.}
	\label{fig1}
\end{figure}

\begin{itemize}
\item The $\nu$LA is intended to approximate the radial neutrino wave function using the $s$-wave component of the orbital angular momentum.
According to Fig.~1, one can observe that
the deviation of the $\nu$LA from the exact numerical solution may distort the spectrum, which will reach around 1\% for large
$Q$ branches, and at the level of 0.1\% for small $Q$ branches. This behaviour can be explained by the property of the spherical Bessel function.
The $s$-wave component at the nuclear surface $j_0(p_\nu R)$ is always smaller than unity and it approaches to unity as $p_\nu$ becomes smaller.
Therefore $\nu$LA always overestimates the spectrum when the $s$-wave component is the dominate one. When $p_\nu$ is large enough and approaching
to the endpoint energy, the $p$-wave component may be sizable and contribute to the additional spectrum distortion near the endpoint energies.
On the other hand, there is no obvious effect for different choices of the nuclear charge $Z$,
which can be simply understood by the neutrality of the neutrino, and its indirect effect is from the nuclear radius when applying the SA.

\item The analytical Fermi function is obtained by taking the leading terms of $\mathcal{O}(\alpha Z)$, $\mathcal{O}(p_{e}R)$ and $\mathcal{O}(m_{e}R)$.
From Fig.~1, we can observe that the Fermi function overestimates the spectral strength in comparison to the exact numerical one,
in particular for the high electron or low antineutrino energy regions. The deviation becomes larger as the electron energy increases
(or equivalently the antineutrino energy decreases). This is reasonable because the Fermi function only includes the leading terms of $p_{e}R$
and the approximation would become
worse when the electron momentum/energy becomes larger. This property is also applicable when we compare the left and right panels of Fig.~1,
where larger endpoint energies $Q$ will have larger spectral deviation. By comparing the upper and lower panels of Fig.~1,
another parameter that may have significant contributions to the spectral deviation is the nuclear charge $Z$,
where the deviation would be more significant for the branches with the larger nuclear charge.
Therefore, it is clear that the spectral deviation of the Fermi function mainly comes from the
endpoint energy $Q$ and the nuclear charge $Z$, which can reach the magnitude of 20\% for the branch with $Q\simeq 8$ MeV and $Z\simeq55$.

\end{itemize}

Second, we want to discuss the results for the uniform charge distribution of the Coulomb potential
and test the validity of the approximation for the finite size effect.
The numerical finite size corrections are accomplished with Eq.~(8) by comparing the numerical results between the uniform and point charge distributions.
The analytical approximations of the finite size effect are calculated using Eqs.~(15)---(17).
Comparisons between the numerical calculations and analytical approximations of four representative decay branches are illustrated in Fig.~(2),
where the analytical finite size effects (blue) are defined as ratios of the analytical Fermi function approximation (FFA) with and without finite size corrections (FS, {see Eq~(\ref{eqFermi}) with and without $\delta_{FS}$}), and the numerical finite size effects (red) are defined as ratios of the numerical spectra for the uniform charge distribution (UD) and
point charge distribution (PD) without taking the $\nu$LA approximation {(see Eq~(\ref{eqSA}))}.
The solid and dashed lines are for the electron and antineutrino spectra respectively.
\begin{figure}[t]
	\centering
	\includegraphics[width=1\linewidth]{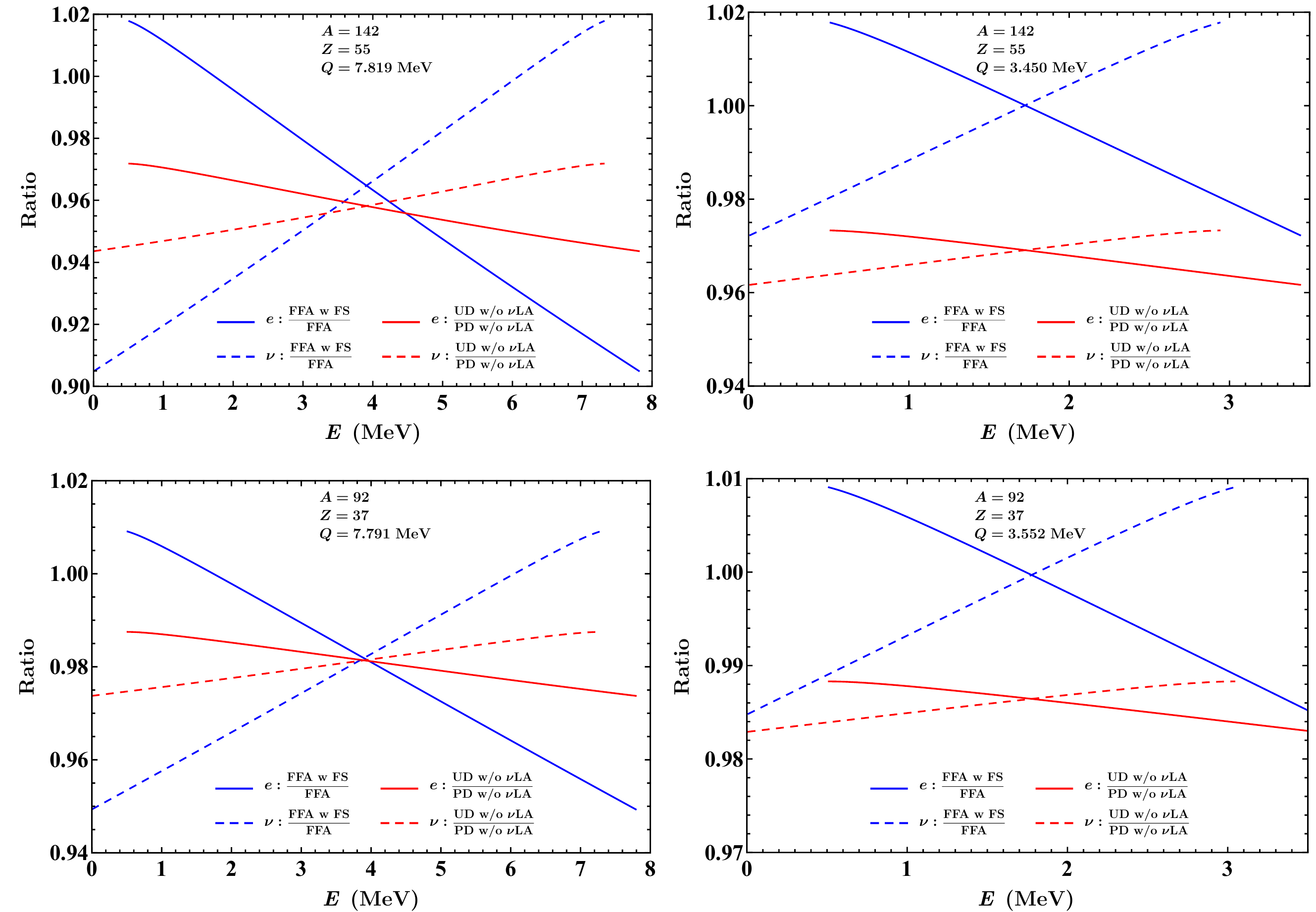}
	\caption{Comparisons between the numerical calculations and analytical approximations of the finite size effect, where the analytical finite size
effects (blue) are defined as ratios of the analytical Fermi Function approximation (FFA) with and without finite size corrections (FS, {see Eq~(\ref{eqFermi}) with and without $\delta_{FS}$}),
and the numerical finite size effects (red) are defined as ratios of the numerical spectra for the uniform charge distribution (UD) and
point charge distribution (PD) without taking the $\nu$LA approximation {(see Eq~(\ref{eqSA}))}.
The solid and dashed lines are for the electron and antineutrino energy spectra respectively.}
	\label{fig2}
\end{figure}

The absolute sizes of both analytical and numerical finite size corrections show reduction of the spectral strength
in most of the energy range, but the size and slope of analytical finite size corrections are much larger than those of the numerical ones.
For the large $Q$ and large $Z$ branch the largest finite size correction will be around 10\% for the analytical calculation,
whereas it is only 3\% for the numerical calculation.
For the relative slope, the analytical and numerical finite size effects have similar behaviour that can be enhance the antineutrino spectra in the high energy part
but reduce the spectra in the low energy part. However they are more significant for the analytical calculations than those of the numerical ones.

From the general consideration of numerical calculations, the analytical Fermi function and finite size correction each has the spectral deviation
of $\mathcal{O}(10\%)$, thus it may be more severe when one considers both effects. However, the situation is the opposite.
Since the total decay magnitude is always normalized to the experimental decay rate, it is thus more transparent to consider the
relative spectral deviations of both analytical and numerical calculations.
The Fermi function shifts more strength to the low antineutrino energies relative to the exact numerical one,
whereas the analytical finite size correction shifts more strength to the high antineutrino energies relative to the numerical finite size correction,
namely, compared to the exact numerical calculations, the spectral deviations induced by the Fermi function and the analytical finite size effect are
in the opposite directions.
\begin{figure}[t]
	\centering
	\includegraphics[width=1\linewidth]{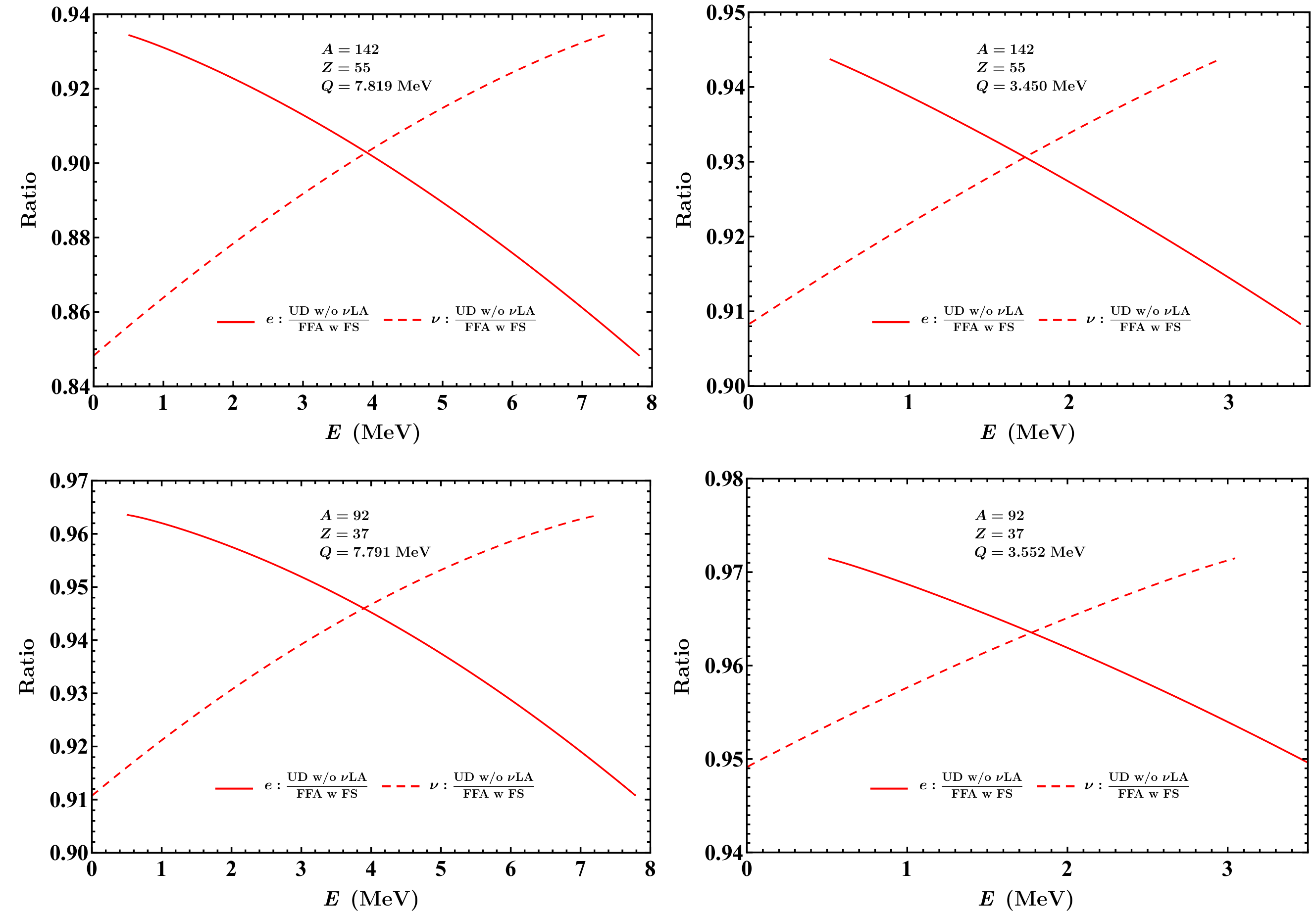}
	\caption{Spectral ratios between the numerical spectra of the uniform charge distribution (UD) without the $\nu$LA {(see Eq~(\ref{eqSA}))}
and the analytical Fermi function {approximation (FFA)} with the finite size effect (FS, {see Eq~(\ref{eqFermi})}) are illustrated.
The solid and dashed lines are for the electron and antineutrino energy spectra respectively.}
	\label{fig3}
\end{figure}
In Fig.~3, the spectral ratios between the numerical spectra of the uniform charge distribution (UD) without the $\nu$LA {(see Eq~(\ref{eqSA}))}
and the analytical Fermi function {approximation (FFA)} with the finite size effect (FS, {see Eq~(\ref{eqFermi})}) are illustrated for the representative branches.
In the upper left panel, the largest spectral deviation would approach 15\% and 6\% for the low and high energy parts of the antineutrino spectra.
For the relative deviation,
it corresponds to a total spectral variation of 9\% for the branch with $Q\simeq 8$ MeV and $Z\simeq55$.

Finally, before finishing this section we would like to discuss the different effects of different extended charge distributions,
where the Gaussian charge distribution (GD) and Fermi charge distributions (FD) are used to compare to the spectra of the uniform charge distribution (UD).
The spectral ratios of different numerical calculations are illustrated in Fig.~4, where one can observe that
the spectral deviations of GD and FD are at most at the levels of {2\textperthousand} and {0.2\textperthousand}, respectively.
{Note that All the spectra are calculated using the SA (see Eq~(\ref{eqSA}))}. Therefore, the effects of different extended charge distributions
are pretty small and negligible, and it would be enough to use the uniform charge distribution in the future calculations of reactor antineutrino spectrum.
\begin{figure}[t]
	\centering
	\includegraphics[width=1\linewidth]{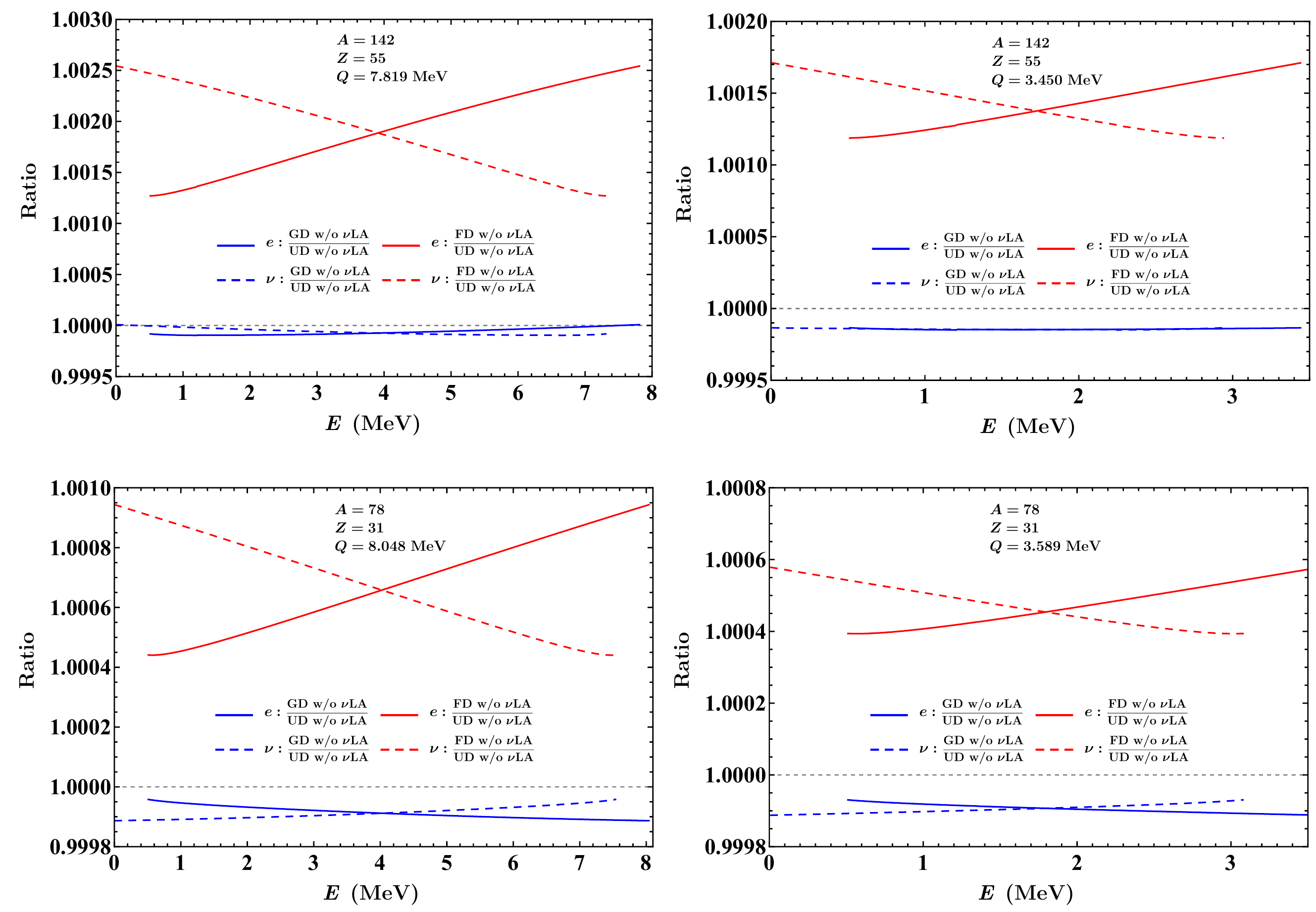}
	\caption{Ratios of absolute spectra numerically calculated using the Gaussian charge distribution (GD) or the Fermi charge distribution (FD)
to those with the uniform charge distribution (UD). {All the spectra are calculated using the SA (see Eq~(\ref{eqSA}))}. The solid and dashed lines are for the electron and antineutrino energy spectra respectively.}
	\label{fig8}
\end{figure}

\section{Reactor antineutrino spectrum}

With the calculation of single beta decay spectrum, we can now obtain the aggregate antineutrino spectrum
associated with the specific fission fuel $k$ ($k= {^{235}{\rm U}}, {^{238}{\rm U}}, {^{239}{\rm Pu}}, {^{241}{\rm Pu}}$),
which is the summation of the contributions of all beta decay branches from the fission fragments, namely,
\begin{eqnarray}
N_{k} =  \sum^{n_f}_{f=1} Y^k_f(Z,A,t) \sum^{n_i}_{i=1} b_{f,i} N_{\nu} (E_{\nu},Q^i,Z) \;,
\end{eqnarray}
where $Y^k_f(Z,A,t)$ is the activity of the $f$-th fission fragment with $A$ and $Z$ at time $t$,
which converges to the cumulative fission yield and is independent of $t$ after sufficient burning time,
$b_{f,i}$ stands for the branching ratios of the transition associated with the endpoint $Q^i$. $E_{\nu}$ is the energy of the emitted antineutrino.
Then the total antineutrino spectrum emitted by a reactor is determined by the summation of the contributions of all four fission fuels:
\begin{eqnarray}
N_{\rm tot} = \sum^{}_{k} \alpha_k N_k \;,
\end{eqnarray}
in which $\alpha_k$ is the fission fraction of the fission fuel $k$.

The spectrum associated with each of the fissionable isotopes $^{235}{\rm U}$, $^{238}{\rm U}$, $^{239}{\rm Pu}$ and $^{241}{\rm Pu}$
consists of more than 6000 beta decay transitions of the fission fragments. Thus in order to obtain the aggregate spectrum, we require not only the normalized single beta decay spectra, but also the corresponding branching ratios and the cumulative fission yields of corresponding fission fragments.
In this work, we employ the latest nuclear database for these relevant nuclear data, where the cumulative fission yield data is taken from the Evaluated Nuclear Data File (ENDF) B-\uppercase\expandafter{\romannumeral8}.0 and the beta decay information is from the database of the Evaluated Nuclear Structure Data Files (ENSDF).

In previous \textit{ab initio} calculations, analytical Fermi function and additional corrections are used to describe single beta decay spectrum.
However, as we have demonstrated in the previous section, these analytical approximations are not always accurate,
and the largest spectral deviation could be at the level of
$\mathcal{O}(10\%)$. Therefore it is necessary to study how they affect the isotopic antineutrino spectrum calculated from Eq.~(17).
In this work, we employ the numerical calculations of the radial electron wave functions in Eq.~(8) with the uniform charge distribution and the SA approximation,
and compare with those using the analytical calculation in Eq.~(17).
The numerical calculations of the radial electron wave functions are obtained from the {\it Radial} package~\cite{Fang:2019qcu,SFW95}.

\begin{figure}
	\centering
	\includegraphics[width=0.85\linewidth]{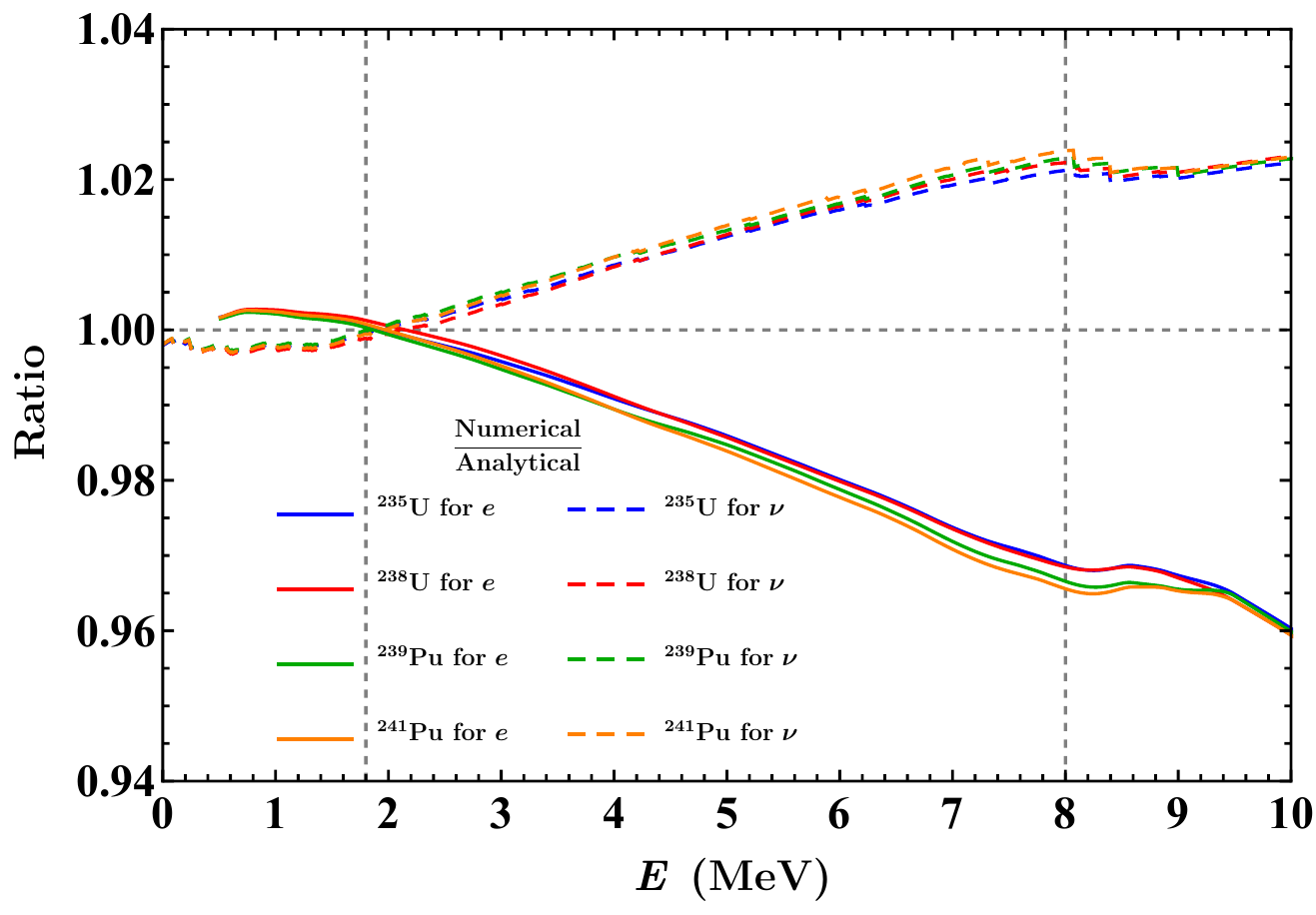}
	\caption{Ratios of the aggregate spectra between the numerical results obtained from exact calculations of the uniform charge distribution to the analytical ones
obtained from approximate calculations of the Fermi function and finite size correction for the fissionable isotopes $^{235}{\rm U}$, $^{238}{\rm U}$, $^{239}{\rm Pu}$ and $^{241}{\rm Pu}$ respectively.}
	\label{fig10}
\end{figure}

In Fig.~5, ratios of the aggregate spectra between the numerical results obtained from exact calculations of the uniform charge distribution to the analytical ones
obtained from approximate calculations of the Fermi function and finite size correction are illustrated for the fissionable isotopes $^{235}{\rm U}$, $^{238}{\rm U}$, $^{239}{\rm Pu}$ and $^{241}{\rm Pu}$.
Compared with the aggregate spectra using the numerical radial electron wave function, the analytical calculation would introduce a reduction of
the low energy antineutrino spectra and an enhancement in the high antineutrino energy spectra.
In the energy range between 2 MeV and 8 MeV, the antineutrino spectral deviations can reach around 2\% to 3\% at $E_\nu = 8 ~{\rm MeV}$
compared to the numerical calculations.
Similar but opposite behavior can be observed for the electron energy spectra, where
the spectral deviations of the analytical calculations are around 3\% to 4\% at $E_{e} = 8 ~{\rm MeV}$ compared to the numerical calculations.
Note that these spectral deviations are much smaller than those of individual beta decay spectra,
which are affected by the small cumulative fission yields, small branching ratios and average effect between thousands of beta decay branches.
Moreover, one can observe that the behavior of spectral variations is rather similar for all the fissionable isotopes and
their differences are within the 0.5\% level, which indicates that if the single beta decay spectrum is really contribute to some levels of
the experimental observed spectral excess, it should be universal for all the four fissionable isotopes.

Finally we would like to comment that the current numerical calculations are still taking the lepton wave functions at the nuclear surface
in order to factorize the nuclear matrix element. In order to go beyond the SA approximation, one needs to know the radial dependence of
the nuclear matrix element, which will
rely on the nuclear many body calculations. Some preliminary studies~\cite{Fang:2019qcu} show that the allowed GT decays are dominated by
the transitions between two single particle orbitals near the nuclear surface, but dedicated study on the nuclear structure information
should be done for the precise beta spectral calculation.
Moreover, even larger spectral deviation from the nuclear structure information has been observed for the first forbidden decays~\cite{Fang:2015cma},
which may have large contributions to the high energy range of the reactor antineutrino spectrum~\cite{Li:2019quv}.
Therefore, further study on these directions would be interesting and constitutes the main goal of our future work.

\section{Concluding Remarks}

The appearance of the reactor flux and spectrum anomalies requires new accurate theoretical predictions for the reactor antineutrinos
associated with the fission isotopes $^{235}{\rm U}$, $^{238}{\rm U}$, $^{239}{\rm Pu}$ and $^{241}{\rm Pu}$.
Many possible issues in the reactor antineutrino flux calculations have been investigated, but none of them can fully account for all these observed anomalies.
In this work, we have proposed a new \textit{ab initio} calculation of the isotopic reactor antineutrino fluxes with the exact calculation of the
radial lepton wave functions, assuming all the decay branches are allowed GT transitions.

With the assumption of the SA approximation, the single beta decay spectrum is numerically calculated using the {\it Radial} package,
and it has been demonstrated that the approximation of the analytical Fermi function and finite size effect each could have the
largest spectral deviation of $\mathcal{O}(10\%)$, whereas the effect of their combination
could result in spectral deviations at the level of 5\%---10\%.
Meanwhile, we also find that several forms of the extended charge distributions have negligible effects on the spectral variation.
Using the state-of-the-art nuclear databases, our new \textit{ab initio} calculations have shown sizable but opposite
spectral deviations at the level of 2\%---4\% in the antineutrino and electron spectra which may partially contribute
to the observed spectral excess in the high energy range. We also find that the effect of analytical beta decay spectrum approximation is rather universal
for all the four fissionable isotopes.
{Finally we want to emphasize that our new \textit{ab initio} calculation based on the numerical wave functions should be refined and go
beyond the SA, even be applied to the first forbidden transitions if nuclear many body calculations can be implemented, inclusion of the latest Pandemonium free data
is also a necessary step forward.
All of these ingredients will be the main goal of our future studies.}

\section*{Acknowledgments}
This work was supported in part by the National Natural Science Foundation of China under Grant No.~11835013, No.~11775231, and No.~12075255,
by Beijing Natural Science Foundation under Grant No.~1192019,
by the CAS Center for Excellence in Particle Physics (CCEPP),
by the CAS "Light of West China" Program and CAS key research program (XDPB09-2),
and by the National Key Research and Development Program (MOST 2016YFA0400501) from the Ministry of Science and Technology of China.


\begin{thebibliography}{99}

\bibitem{Tanabashi:2018oca}
  M.~Tanabashi {\it et al.} [Particle Data Group],
  {\it Review of Particle Physics},
  Phys.\ Rev.\ D {\bf 98}, no. 3, 030001 (2018).

%
%
%
%
%
%
%

\bibitem{Hayes:2016qnu}
  A.~C.~Hayes and P.~Vogel,
  {\it Reactor Neutrino Spectra},
  Ann.\ Rev.\ Nucl.\ Part.\ Sci.\  {\bf 66}, 219 (2016).

\bibitem{Davis:1979gg}
  B.~R.~Davis, P.~Vogel, F.~M.~Mann and R.~E.~Schenter,
  {\it Reactor Anti-neutrino Spectra And Their Application To Anti-neutrino Induced Reactions},
  Phys.\ Rev.\ C {\bf 19}, 2259 (1979).

\bibitem{Avignone:1980qg}
  F.~T.~Avignone, III and C.~D.~Greenwood,
  {\it Calculated spectra of antineutrinos from U235, U238, and Pu239, and antineutrino-induced reactions},
  Phys.\ Rev.\ C {\bf 22}, 594 (1980).

\bibitem{Klapdor:1982zz}
  H.~V.~Klapdor and J.~Metzinger,
  {\it Antineutrino Spectrum from the Fission Products of Pu-239},
  Phys.\ Rev.\ Lett.\  {\bf 48}, 127 (1982).

\bibitem{Klapdor:1982sf}
  H.~V.~Klapdor and J.~Metzinger,
  {\it Calculation Of The Anti-neutrinos Spectrum From Thermal Fission Of U-235},
  Phys.\ Lett.\ B {\bf 112}, 22 (1982).

\bibitem{Mueller:2011nm}
  T.~A.~Mueller {\it et al.},
  {\it Improved Predictions of Reactor Antineutrino Spectra},
  Phys.\ Rev.\ C {\bf 83}, 054615 (2011).

\bibitem{Fallot:2012jv}
  M.~Fallot {\it et al.},
  {\it New antineutrino energy spectra predictions from the summation of beta decay branches of the fission products},
  Phys.\ Rev.\ Lett.\  {\bf 109}, 202504 (2012).

\bibitem{Hayes:2013wra}
A.~C.~Hayes, J.~L.~Friar, G.~T.~Garvey, G.~Jungman and G.~Jonkmans,
 {\it Systematic Uncertainties in the Analysis of the Reactor Neutrino Anomaly},
Phys.\ Rev.\ Lett.\  {\bf 112}, 202501 (2014).

\bibitem{Dwyer:2014eka}
  D.~A.~Dwyer and T.~J.~Langford,
  {\it Spectral Structure of Electron Antineutrinos from Nuclear Reactors},
  Phys.\ Rev.\ Lett.\  {\bf 114}, no. 1, 012502 (2015).

\bibitem{Hayes:2015yka}
  A.~C.~Hayes, J.~L.~Friar, G.~T.~Garvey, D.~Ibeling, G.~Jungman, T.~Kawano and R.~W.~Mills,
  {\it Possible origins and implications of the shoulder in reactor neutrino spectra},
  Phys.\ Rev.\ D {\bf 92}, no. 3, 033015 (2015).

\bibitem{Littlejohn:2018hqm}
  B.~R.~Littlejohn, A.~Conant, D.~A.~Dwyer, A.~Erickson, I.~Gustafson and K.~Hermanek,
  {\it Impact of Fission Neutron Energies on Reactor Antineutrino Spectra},
  Phys.\ Rev.\ D {\bf 97}, no. 7, 073007 (2018).

\bibitem{Yoshida:2018zga}
  T.~Yoshida, T.~Tachibana, S.~Okumura and S.~Chiba,
  {\it Spectral anomaly of reactor antineutrinos based on theoretical energy spectra},
  Phys.\ Rev.\ C {\bf 98}, no. 4, 041303 (2018).

\bibitem{Estienne:2019ujo}
  M.~Estienne {\it et al.},
  {\it Updated Summation Model: An Improved Agreement with the Daya Bay Antineutrino Fluxes},
  Phys.\ Rev.\ Lett.\  {\bf 123}, no. 2, 022502 (2019).
%
\bibitem{VonFeilitzsch:1982jw}
  F.~Von Feilitzsch, A.~A.~Hahn and K.~Schreckenbach,
  {\it Experimental Beta Spectra From Pu-239 And U-235 Thermal Neutron Fission Products And Their Correlated Anti-neutrinos Spectra},
  Phys.\ Lett.\  {\bf 118B}, 162 (1982).

\bibitem{Schreckenbach:1985ep}
  K.~Schreckenbach, G.~Colvin, W.~Gelletly and F.~Von Feilitzsch,
  {\it Determination Of The Anti-neutrino Spectrum From U-235 Thermal Neutron Fission Products Up To 9.5-mev},
  Phys.\ Lett.\  {\bf 160B}, 325 (1985).

\bibitem{Hahn:1989zr}
  A.~A.~Hahn, K.~Schreckenbach, G.~Colvin, B.~Krusche, W.~Gelletly and F.~Von Feilitzsch,
  {\it Anti-neutrino Spectra From $^{241}$Pu and $^{239}$Pu Thermal Neutron Fission Products},
  Phys.\ Lett.\ B {\bf 218}, 365 (1989).

\bibitem{Haag:2013raa}
  N.~Haag, A.~G\"{u}tlein, M.~Hofmann, L.~Oberauer, W.~Potzel, K.~Schreckenbach and F.~M.~Wagner,
  {\it Experimental Determination of the Antineutrino Spectrum of the Fission Products of $^{238}$U},
  Phys.\ Rev.\ Lett.\  {\bf 112}, no. 12, 122501 (2014).

\bibitem{Vogel:2007du}
  P.~Vogel,
  {\it Conversion of electron spectrum associated with fission into the antineutrino spectrum},
  Phys.\ Rev.\ C {\bf 76}, 025504 (2007).

\bibitem{Huber:2011wv}
  P.~Huber,
  {\it On the determination of anti-neutrino spectra from nuclear reactors},
  Phys.\ Rev.\ C {\bf 84}, 024617 (2011)
  Erratum: [Phys.\ Rev.\ C {\bf 85}, 029901 (2012)].

\bibitem{Hayen:2019ieh}
  L.~Hayen, J.~Kostensalo, N.~Severijns and J.~Suhonen,
  {\it First-forbidden transitions in reactor antineutrino spectra},
  Phys.\ Rev.\ C {\bf 99}, no. 3, 031301 (2019).

\bibitem{Li:2019quv}
  Y.~F.~Li and D.~Zhang,
  {\it New Realization of the Conversion Calculation for Reactor Antineutrino Fluxes},
  Phys.\ Rev.\ D {\bf 100}, no. 5, 053005 (2019).

\bibitem{Hayen:2019eop}
  L.~Hayen, J.~Kostensalo, N.~Severijns and J.~Suhonen,
  {\it First-forbidden transitions in the reactor anomaly},
  arXiv:1908.08302 [nucl-th].

\bibitem{Mention:2011rk}
  G.~Mention, M.~Fechner, T.~Lasserre, T.~A.~Mueller, D.~Lhuillier, M.~Cribier and A.~Letourneau,
  {\it The Reactor Antineutrino Anomaly},
  Phys.\ Rev.\ D {\bf 83}, 073006 (2011).

\bibitem{Giunti:2017yid}
  C.~Giunti, X.~P.~Ji, M.~Laveder, Y.~F.~Li and B.~R.~Littlejohn,
  {\it Reactor Fuel Fraction Information on the Antineutrino Anomaly},
  JHEP {\bf 1710}, 143 (2017)

\bibitem{Giunti:2019qlt}
  C.~Giunti, Y.~F.~Li, B.~R.~Littlejohn and P.~T.~Surukuchi,
 {\it Diagnosing the Reactor Antineutrino Anomaly with Global Antineutrino Flux Data},
  Phys.\ Rev.\ D {\bf 99}, no. 7, 073005 (2019).

\bibitem{Berryman:2019hme}
  J.~M.~Berryman and P.~Huber,
 {\it Reevaluating Reactor Antineutrino Anomalies with Updated Flux Predictions},
  arXiv:1909.09267 [hep-ph].

\bibitem{Giunti:2019fcj}
  C.~Giunti, Y.~F.~Li and Y.~Y.~Zhang,
 {\it KATRIN bound on 3+1 active-sterile neutrino mixing and the reactor antineutrino anomaly},
  arXiv:1912.12956 [hep-ph].

\bibitem{Seo:2014xei}
  S.~H.~Seo [RENO Collaboration],
  {\it New Results from RENO and The 5 MeV Excess},
  AIP Conf.\ Proc.\  {\bf 1666}, 080002 (2015).

\bibitem{An:2015nua}
  F.~P.~An {\it et al.} [Daya Bay Collaboration],
  {\it Measurement of the Reactor Antineutrino Flux and Spectrum at Daya Bay},
  Phys.\ Rev.\ Lett.\  {\bf 116}, no. 6, 061801 (2016)
  Erratum: [Phys.\ Rev.\ Lett.\  {\bf 118}, no. 9, 099902 (2017)].

\bibitem{Adey:2019ywk}
  D.~Adey {\it et al.} [Daya Bay Collaboration],
  {\it Extraction of the $^{235}$U and $^{239}$Pu Antineutrino Spectra at Daya Bay},
  Phys.\ Rev.\ Lett.\  {\bf 123}, no. 11, 111801 (2019)
  [arXiv:1904.07812 [hep-ex]].

\bibitem{Abe:2015rcp}
  Y.~Abe {\it et al.} [Double Chooz Collaboration],
  {\it Measurement of $\theta_{13}$ in Double Chooz using neutron captures on hydrogen with novel background rejection techniques},
  JHEP {\bf 1601}, 163 (2016).

\bibitem{Ko:2016owz}
  Y.~J.~Ko {\it et al.} [NEOS Collaboration],
  {\it Sterile Neutrino Search at the NEOS Experiment},
  Phys.\ Rev.\ Lett.\  {\bf 118}, no. 12, 121802 (2017).

\bibitem{Hardy:1977suw}
  J.~C.~Hardy, L.~C.~Carraz, B.~Jonson and P.~G.~Hansen,
   {\it The essential decay of pandemonium: A demonstration of errors in complex beta-decay schemes},
  Phys.\ Lett.\ B {\bf 71}, 307 (1977).

\bibitem{BB-Book}
H.~Behrens and W.~B\"{u}hring,
  {\it Electron Radial Wave Functions and Nuclear Beta Decay}, (Clarendon, Oxford, 1982).

\bibitem{Fang:2019qcu}
  D.~L.~Fang,
  {\it Allowed $\beta$-decay spectrum with numerical electron wave functions},
  arXiv:1907.04560 [nucl-th].

%
%
%
%
%
%
%
%
%
%
%
%
%
%
%
%
%


\bibitem{Fermi:1934hr}
  E.~Fermi,
  {\it An attempt of a theory of beta radiation. 1.},
  Z.\ Phys.\  {\bf 88}, 161 (1934).

\bibitem{Wei61}
  H.~A.~Weidenmuller,
  {\it First-Forbidden Beta Decay},
  Rev.\ Mod.\ Phys.\  {\bf 33}, 574 (1961).

\bibitem{BBC77}
W.~Bambynek {\it et al.},
  {\it Orbital electron capture by the nucleus},
  Rev.\ Mod.\ Phys.\  {\bf 49}, 77 (1977)
  Erratum: [Rev.\ Mod.\ Phys.\  {\bf 49}, 961 (1977)].

\bibitem{Elton}
L.~R.~B.~Elton,
{\it A semi-empirical formula for the nuclear radius},
Nucl.\ Phys.\  {\bf 5}, 173 (1958).

\bibitem{SFW95}
F.~Salvat, J.~Fernadez-Varea and W.~Williamson,
{\it Accurate numerical solution of the radial Schr\"{o}dinger and Dirac wave equations},
Comput.\ Phys.\ Commun.\ {\bf 90}, 151(1995).

\bibitem{Wang:2016rqh}
X.~B.~Wang, J.~L.~Friar and A.~C.~Hayes,
{\it Nuclear Zemach moments and finite-size corrections to allowed $\beta$ decay},
Phys.\ Rev.\ C {\bf 94}, 034314 (2016).

\bibitem{Wilkinson}
D.~H.~Wilkinson,
{\it Evaluation of beta-decay: II. Finite mass and size effects},
Nucl.\ Phys.\ Instrum.\ Methods Phys.\ Res.,\ Sect.\ A {\bf 290}, 509 (1990).

\bibitem{Fang:2015cma}
  D.~L.~Fang and B.~A.~Brown,
  {\it Effect of first forbidden decays on the shape of neutrino spectra},
  Phys.\ Rev.\ C {\bf 91}, no. 2, 025503 (2015)
  Erratum: [Phys.\ Rev.\ C {\bf 93}, no. 4, 049903 (2016)]


\end{thebibliography}
\end{document}